# Anisotropic surface plasmon-coupled emission from structured PMMA films doped with Rhodamine B

D.G.Zhang[1], X.-C.Yuan[2a], G.H.Yuan[1], K.J.Moh[1], P. Wang[3], H. Ming[3]

[1]*School of Electrical and Electronic Engineering, Nanyang Technological University, Nanyang Avenue, 639798, Singapore*

[2]*Institute of Modern Optics, Key Laboratory of Optoelectronic Information Science & Technology, Ministry of Education of China, Nankai University, Tianjin 300071, People's Republic of China*

[3]*Institute of Photonics, University of Science and Technology of China, Hefei, Anhui, 230026, People's Republic of China*

---

[a] Electronic mail: xcyuan@nankai.edu.cn

We report observation of anisotropic surface plasmon-coupled emission (SPCE) from structured PMMA films doped with Rhodamine B (RhB) and deposited on silver films. For a given structure, it is observed that relative position of incident laser spot to the structured PMMA and polarization properties of the laser beam can be used to modulate the SPCE patterns. The proposed method enables azimuthally controlling directional SPCE in a deterministic manner.

Surface plasmons polaritons (SPPs) are light waves trapped at the metal/dielectric interface owing to its resonant interaction with conduction electrons at the metal surface[1, 2]. There is growing interest in the interactions between fluorophores and SPPs, where spectra property, emitting direction and lifetime of the fluorescence will be modified due to the presence of SPPs[3-9]. A particular area of investigation is focused on the phenomenon of surface plasmon-coupled emission (SPCE) [10-13], which occurs due to localization of fluorophores near a thin silver film on a transparent substrate. Radiation from fluorophores transfers to SPPs modes supported by the silver film and then enters into the substrate at the surface plasmon resonance (SPR) angle (radial angle). The SPCE displays strong directional emission and unique polarization (p-polarized for all the points on the SPCE ring) properties with potentials in sensing applications. In previous studies, the SPCE intensities are distributed in circular rings uniformly, where fluorophores are emitted at the SPR angle (radial angle) in all azimuthal directions. In these experiments, the fluorophores are doped in an extended thin film with random orientations[10-14]. For directionally oriented fluorescence molecules, in contrast, the SPCE pattern exhibited a non-uniform anisotropic intensity distribution [13, 15, 16]. In this Letter, we report observation of an anisotropic SPCE azimuthally pattern by structured dielectric films although the doped fluorescence molecules are randomly oriented. For a given shape, the intensity distribution on SPCE ring can be modulated by polarization of the excitation beam or relative irradiated position to the structure.

The procedure of samples preparation is described below. Rhodamine (RhB) molecules (0.1mg/ml) were dissolved in PMMA solution (950K PMMA, Solids: 2% in Anisole, from MICRO.CHEM corp.) for about 48 hours. The doped solution was agitated by an ultrasonic disrupter for 30 minutes. Then the mixed solution was spin-coated onto a 45 nm thick silver film to get a uniform film of about 80nm thickness. The solvent was removed by put the film on hotplate for 10 min at 105°C. Structured PMMA films on silver films were inscribed by Electron Beam Lithography (Raith GmbH, e_LiNe) technique and standard developing procedures.

Leakage radiation microscopy (LRM), a commonly used technique in SPR experiments[17-19], was used to characterize the SPCE by observing fluorescence images in real and reciprocal spaces respectively. A laser beam with 532 nm wavelength was tightly focused by an oil-immersion objective (60X, numerical aperture (N.A.), 1.42) to excite the RhB molecules. Details about the experimental setup can be found in reference [20].

Figure 1 (a) and (b) depict the calculated two-dimensional intensity distributions of the tightly focused linearly and circularly polarized 532 nm beams at the focal plane respectively [21, 22]. In the calculation, refractive index of the glass substrate, silver, RhB doped PMMA films and airs are 1.516, 0.129+i 3.193, 1.49 and 1 respectively[23]. Thickness of Ag and PMMA films are 45 nm and 80 nm respectively. Intensity distribution of the focused circularly polarized beam is circularly symmetric and strong in the center. In the case of linear polarization, the in-plane field intensity is strong close to the polarization direction (horizontal axis) and diminishes away from the horizontal axis. The field can experimentally be imaged in real time by detecting fluorescence of the molecules doped in the PMMA films[24]. Figure 1 (c) and (d) give the direct space fluorescence images under irradiation by the focused linearly and circularly polarized beams respectively. It shows that intensity distributions of the experimental results are consistent with the calculations as shown in Figure 1 (a) and (b).

The circularly polarized laser beam was initially used as the excitation source due to its circular symmetry as shown in Figure 1 (b) and (d). A triangularly shaped PMMA film was selected at first to demonstrate the anisotropic SPCE. Figure 2 (a) gives the bright field transmission image acquired by a CCD camera without a long pass edge filter to reject the exciting beam. It is noted that the laser beam was attenuated greatly so that the focal point of the excitation beam can be imaged onto the CCD. Side length of the equilateral triangle is 10 µm. Figure 2 (b) shows the direct space fluorescence image with the laser power enhanced to about 0.2 mW. It shows that there are attenuated waves propagating on the bare Ag film near three sides of the triangle. Since the fluorescence molecules were only doped in the PMMA film, the attenuated waves can only be attributed to the SPPs-1 wave on the interface of Air/Ag generated by the excited fluorescence molecules. The interference fringes of the SPPs waves-1(Air/Ag interface) on Figure 2 (b) is due to coherence of the excitation laser (continue wave) and multi-reflections on optical elements, which also appear in the images of SPPs captured by the LRM with a coherent laser source[17]. The SPPs-1 waves marked with 1, 2, and 3 propagate mainly along the direction perpendicular to triangle side and attenuate as propagated. The SPPs-2 waves propagating on the Air/PMMA/Ag interface cannot be clearly imaged because the excitation area is also the propagating area, and it is difficult to distinguish the excitation source and the propagating waves[18]. In this Letter, we mainly investigate the behaviors of SPCE related to the SPPs-1 waves.

Figure 2 (c) demonstrates the corresponding Fourier plane fluorescence image which gives out wave-vector content of the emission. It shows the commonly known SPCE rings: the inner ring is caused by the SPPs-1 waves and the outer by the SPPs-2 waves. The two rings indicate that the fluorescence mainly emitted in two angles (radial angle). Based on the known N.A of the oil-immersed objective, the emitting angles can be estimated as about $43^o$ and $69^o$, corresponding to the calculated SPR angles ($43.6^o$, $69.4^o$, n= 0.12+3.547i for Ag) at the peak wavelength of RhB fluorescence (576 nm). The white boxes labeled with 1, 2, 3 are three brighter arcs on the inner ring, indicating that the SPCE intensities are anisotropic azimuthally and they are stronger in these three areas. This is consistent with the findings in Figure 2 (b), namely the propagation and intensity distribution of the SPPs-1 imaged in the real space are correspondingly reflected in the reciprocal space, such as the 1, 2, and 3 areas marked in Figure 2 (b) and (c). Consequently we conclude that anisotropic propagation of the SPPs-1 wave on the bare Ag film can be realized at the same time as the anisotropic SPCE.

The anisotropic SPCE patterns can also be modified by using circularly polarized laser beam focused onto different positions of a structured PMMA film. Figure 3(a) shows the bright field transmission images focused at one corner of a square shaped PMMA. Figure 3 (b) and (c) are the corresponding direct space and Fourier plane fluorescence images respectively. The excited SPPs-1 waves propagate onto the air side and diverge away from the corner. The areas marked with 1 and 2 are the two stronger SPPs-1 waves with different propagating directions, corresponding to the two brighter arcs marked with white boxes (1 and 2) in Figure 3 (c) respectively. In the direct space image, intensity of the area marked with 2 is found stronger than the mark 1, which was also observed in the Fourier plane image. Figure 3 (d), (e) and (f) explore another case with the focal point at the center of the square side. We observe that the SPPs-1 waves are the strongest in the mark 3 and propagate mainly along the direction perpendicular to the square side as shown in Figure 3 (e). In Figure 3 (f), the strongest arc is marked with 3, corresponding to the strongest area as marked 3 in Figure 3 (e). Figure 3 demonstrates that for the given shape and polarization, different irradiated positions give out different anisotropic SPCE and SPPs-1 wave propagating patterns.

At last, we investigate the influence of polarization. As shown in Figure 2 and Figure 3, intensity distributions of the SPCE ring also clearly represent in the direct space image, so we only give out the direct space images in Figure 4 with different polarization states. The excitation beam is focused onto the centre of the same square as used in Figure 3. The excitation laser beam in (a, b), (c) and (d) are circularly polarized, vertically and horizontally linearly-polarized respectively. Figure 4 (a) shows that the intensity distribution of the SPPs-1 waves at the four sides of the square are the same under irradiation of circularly polarized beam. The propagating directions of the SPPs-1 waves are mainly perpendicular to the four square side, which induce the anisotropic SPCE pattern. This is due to the shape effect of the same mechanism as shown in Figure 2. The SPPs propagating pattern in Figure 4 (a) are different from Figure 2 (b), indicating that different shapes can produce different SPCE pattern. It should be noted, if the circularly polarized beam was focused onto the center of a circular shape PMMA film as shown in Figure 4 (b), that the SPCE pattern will be isotropic due to the symmetry of both excitation sources and the structures. Figure 4 (c) and (d) display that the SPPs-1 waves are stronger along the direction of polarization. The intensity distribution and propagating patterns of the SPPs-1 waves along the polarized direction are different from the perpendicular direction, which can be attributed to the different intensity distributions of the focused linearly and circularly polarized beams in Figure 1. This will induce further anisotropic SPCE except for the shape effect.

In summary, anisotropic SPCE and propagation of SPPs on silver film are realized by using structured PMMA films doped with RhB molecules. The SPCE from oriented fluorescence molecules exhibit an anisotropic pattern with the strongest intensity at azimuthal angles where molecules are oriented. Since it is convenient to fabricate PMMA structures by lithography, various intensity patterns on SPCE ring can be realized by different shapes of the PMMA structure, polarization states and relatively irradiated positions in a deterministic manner. The method provides a control of fluorescence emitting at azimuthal angles in addition to the radial angles in common SPCE.

This work was partially supported by the National Natural Science Foundation of China under Grant No. (10974101), Ministry of Science and Technology of China under Grant no.2009DFA52300 for China-Singapore collaborations and National Research Foundation of Singapore under Grant No. NRF-G-CRP 2007-01. HM acknowledges the funding support given by the National Natural Science Foundation of China under Grant No.60736037 and the National Key Basic Research Program of China under Grant No. 2006CB302905.

**FIGURE CAPTIONS:**

Figure.1: Calculated two-dimensional total field intensity distribution at the focal plane for linearly (a) and circularly (b) polarized 532nm laser beam.

Figure 2: (a): Bright-field transmission image of triangular PMMA film acquired by the CCD camera without long pass filter. The central bright spot is the focused laser spot. (b) direct-space fluorescence image; (c) Fourier plane fluorescence image.

Figure 3: (a) and (d): Bright-field transmission image of square shape PMMA film acquired by the CCD camera without long pass edge filter. The central bright point is the focused laser spot. (b) and (e) are direct-space fluorescence images; (c) and (f) are the Fourier plane fluorescence images.

Figure 4: Direct space fluorescence image with different polarization of the incident laser beam. (a) and (b): the square shape and circular shape PMMA films under irradiation of circularly polarized beam; (c): vertically linear polarization (d): horizontally linear polarization in the case of square shape PMMA film.

**FIGURES:**

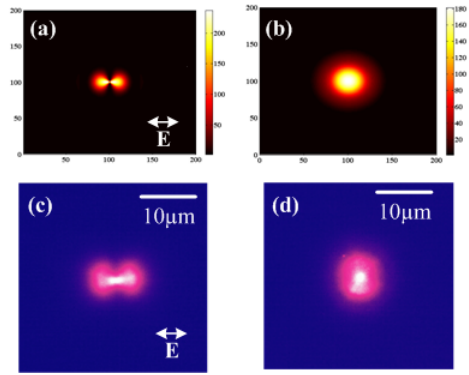

Figure 1

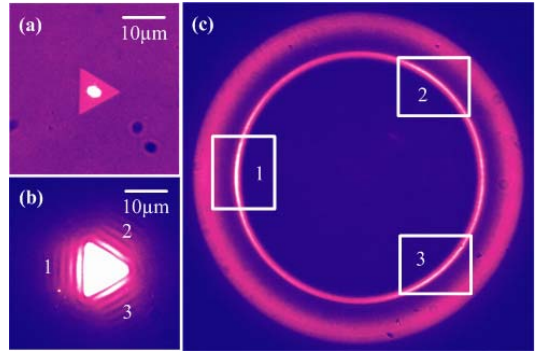

Figure 2

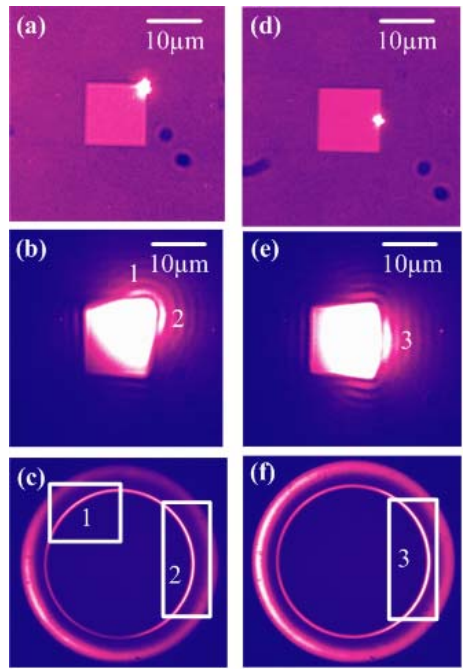

Figure 3

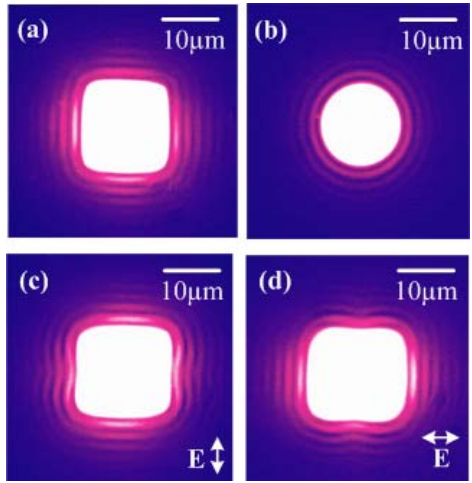

Figure 4